\documentclass[
    aps,
    pra,
    reprint,
    amsmath,
    amssymb,
    superscriptaddress
]{revtex4-2}

\usepackage{graphicx}
\usepackage{bm}
\usepackage{braket}
\usepackage{amsthm}
\usepackage{hyperref}

\newcommand{\avg}[1]{\left\langle #1 \right\rangle}
\newcommand{\ee}{\mathrm{e}}
\newcommand{\ii}{\mathrm{i}}
\newcommand{\dd}{\mathop{}\!\mathrm{d}}
\newcommand{\bs}{\boldsymbol{s}}
\newcommand{\bnab}{\boldsymbol{\nabla}}
\newcommand{\EE}{\mathbb{E}}

\newtheorem*{theorem}{Theorem}

\begin{document}

\title{Dicke superradiance without entanglement: an exact spin-coherent-state diffusion}

\author{Darius Bostan}
\affiliation{CESQ, ISIS, Universit\'e de Strasbourg,
CNRS, 67000 Strasbourg, France}
\affiliation{Department of Physics and Astronomy ``Galileo Galilei'',
University of Padova, 35131 Padova, Italy}

\begin{abstract}
We give an alternative, constructive proof that, starting from the fully excited state, the atomic density matrix of the Dicke superradiance model is, at all times and for an arbitrary number of atoms $N=2j$, a statistical mixture of spin-$j$ coherent states, each a product of identical spin\nobreakdash-$1/2$ coherent states: all atoms share the same single-atom state. This provides an explicit, fully separable decomposition and shows that the Dicke cascade does not generate entanglement among the atoms. The proof builds on an exact phase-space description of superradiance, in which the Lindblad equation is mapped onto a second-order equation for the Glauber--Sudarshan $P$ function over spin coherent states. Owing to the phase invariance of the problem, this equation reduces to a one-dimensional Fokker--Planck equation with non-negative diffusion, which ensures that the phase-averaged $P$ function remains a genuine probability distribution throughout the decay.
\end{abstract}

\maketitle

\section{Introduction}\label{sec:intro}

Since Dicke's original work~\cite{Dicke1954}, superradiance has been a paradigm of collective light--matter interaction. When $N$ identical two-level emitters are confined to a volume much smaller than the cube of the emission wavelength (Dicke limit), they couple identically to the radiation field and radiate cooperatively. Starting from the fully excited state, they emit a burst whose peak intensity scales as $N^2$ rather than $N$, after a delay of order $\ln N/(N\gamma_0)$, where $\gamma_0$ is the single-atom decay rate~\cite{GrossHaroche}.

In this limit the dynamics is that of a collective spin $j=N/2$ decaying down the ladder of Dicke states $\ket{j,m}$. All Dicke states except the two extremal ones are genuinely multipartite entangled~\cite{Huber2011,Lucke2014}. The state of the atoms during the decay, however, is not a single Dicke state but a statistical mixture of them, and a mixture of entangled states can be separable. Wolfe and Yelin certified separability numerically for up to eight atoms~\cite{WolfeYelin2014}. More recently, Rosario \textit{et al.}\ decomposed the decaying state into positive mixtures of spin coherent states (SCS), analytically for two atoms and numerically for up to fifty, and used them to construct low-entanglement quantum-trajectory unravelings~\cite{Rosario2025}. Independently, Bassler proved analytically that the state remains separable at all times for any $N$, by mapping the problem onto a truncated Hausdorff moment problem~\cite{Bassler2025}.

In this work we give an alternative, constructive proof of this result based on phase-space methods~\cite{CahillGlauber,CahillGlauberII,Polkovnikov,RundleEveritt}. These methods map the master equation onto an evolution equation for a quasiprobability distribution. When this equation takes the Fokker--Planck form with a positive-semidefinite diffusion matrix, it is equivalent to a stochastic differential equation whose trajectories sample the dynamics~\cite{Carmichael1999}. For interacting many-body systems, however, this equation generally contains higher-order derivatives or a diffusion matrix without a definite sign, and stochastic sampling usually requires approximations such as the truncated Wigner approximation~\cite{Polkovnikov,Mink2023}.

Here we show that the Dicke cascade is an exception. Our strategy is to work with the $P$ function over SCS~\cite{Arecchi1972,TakahashiShibata1976}, for which the master equation maps exactly onto a second-order equation. Its diffusion matrix is not positive semidefinite, but since both the initial state and the dynamics are invariant under rotations about the $z$ axis, $P$ can be replaced by its azimuthal average, which represents the same state. The averaged dynamics closes on a single variable, the excitation probability per atom $z=\cos^2(\theta/2)$. Up to a change of variables, the resulting one-dimensional equation is the exact diffusion equation of Narducci \textit{et al.}~\cite{Narducci1974}. Its diffusion is non-negative, so $z$ obeys a stochastic differential equation, and the first $N$ moments of $z$ determine exactly the Dicke populations for any finite $N$. The phase-averaged $P$ function therefore remains a genuine probability distribution at all times, and the stochastic process provides an explicit, dynamically generated decomposition of the full atomic state into SCS, which are themselves products of identical single-atom states.

The paper is organized as follows. Section~\ref{sec:model} introduces the model and shows that a non-negative $P$ function implies full separability. Section~\ref{sec:theorem} proves the main result. Section~\ref{sec:discussion} derives the stochastic differential equation from the phase-space description and relates the moments of $z$ to the Dicke populations. Section~\ref{sec:conclusions} concludes.

\section{Model and separability condition}\label{sec:model}

Our goal is to determine whether the joint state of the $N$ atoms during the cascade is a convex combination of product states. We first use the symmetries of the master equation to reduce the dynamics to the Dicke states $\ket{j,m}$ and then show that a non-negative $P$ function over SCS implies separability.

\subsection{Master equation}

We consider $N$ identical two-level atoms with excited state $\ket{e}$ and ground state $\ket{g}$, all initially excited,
\begin{equation}
    \hat\rho(0)=\left(\ket{e}\!\bra{e}\right)^{\otimes N}.
    \label{eq:initial_state}
\end{equation}
In the Dicke limit all atoms couple identically to the radiation field, which therefore interacts with them only through their collective dipole. It is then natural to introduce the collective spin operators
\begin{equation}
    \hat S_- = \sum_{i=1}^{N}\hat\sigma_-^{(i)}, \qquad
    \hat S_z = \frac{1}{2}\sum_{i=1}^{N}\hat\sigma_z^{(i)},
    \label{eq:collective_ops}
\end{equation}
where $\hat\sigma_-^{(i)}=\ket{g}_i\!\bra{e}$ and $\hat\sigma_z^{(i)}=\ket{e}_i\!\bra{e}-\ket{g}_i\!\bra{g}$ act on atom $i$. We further define $\hat S_+=\hat S_-^\dagger$ and $\hat S_{x,y}$ through $\hat S_\pm=\hat S_x\pm\ii\hat S_y$. Tracing out the electromagnetic field in the Born--Markov and rotating-wave approximations~\cite{Lehmberg,Agarwal} yields the Lindblad master equation for the atoms
\begin{equation}
    \frac{\dd\hat\rho}{\dd\tau} = \mathcal{L}[\hat\rho]
    = \hat S_-\hat\rho\,\hat S_+ - \frac{1}{2}\left\{\hat S_+\hat S_-,\hat\rho\right\},
    \label{eq:master}
\end{equation}
where $\tau=\gamma_0 t$ is the dimensionless time. The single jump operator $\hat S_-$ describes the emission of a photon by the ensemble as a whole.

\subsection{Symmetries of the model}

Two symmetries of Eq.~\eqref{eq:master} drastically simplify the problem. First, the total spin $\hat S^2=\hat S_x^2+\hat S_y^2+\hat S_z^2$ commutes with $\hat S_\pm$, so the dynamics does not couple sectors with different total spin. The initial state~\eqref{eq:initial_state} is $\ket{j,j}$ with $j=N/2$, hence the state remains at all times in the $(N+1)$-dimensional symmetric subspace. This subspace is spanned by the Dicke states $\ket{j,m}\propto\hat S_-^{\,j-m}\ket{j,j}$, with $m=-j,\dots,j$, where $\ket{j,m}$ is the symmetric superposition of all configurations with $j+m$ excited atoms.

Second, the dynamics is invariant under rotations about the $z$ axis. With $\hat U_\phi=\ee^{-\ii\phi\hat S_z}$ one has $\hat U_\phi\hat S_-\hat U_\phi^\dagger=\ee^{\ii\phi}\hat S_-$, and since the dissipator in Eq.~\eqref{eq:master} is invariant under a global phase of the jump operator,
\begin{equation}
    \mathcal{L}\bigl[\hat U_\phi\hat\rho\,\hat U_\phi^\dagger\bigr]
    =\hat U_\phi\,\mathcal{L}[\hat\rho]\,\hat U_\phi^\dagger .
    \label{eq:covariance}
\end{equation}
The initial state is invariant under $\hat U_\phi$, hence so is $\hat\rho(\tau)$ at all times, i.e., $[\hat\rho(\tau),\hat S_z]=0$. The density matrix is therefore diagonal in the Dicke basis,
\begin{equation}
    \hat\rho(\tau)=\sum_{m=-j}^{j}p_m(\tau)\ket{j,m}\!\bra{j,m}.
    \label{eq:dicke_diagonal}
\end{equation}
Since the Dicke states with $|m|<j$ are entangled, this form does not reveal whether $\hat\rho(\tau)$ is separable. We therefore expand $\hat\rho(\tau)$ over SCS, which are product states.

\subsection{Spin coherent states and separability condition}

We use the SCS of Ref.~\cite{Arecchi1972}, defined here as rotations of the fully excited Dicke state,
\begin{equation}
\begin{aligned}
    \ket{\Omega} &\equiv \ket{\theta,\phi}=\hat{U}_\phi\, \hat{U}_\theta \ket{j,j},\\
    \hat{U}_\phi &= \ee^{-\ii\phi\hat S_z}, \qquad \hat{U}_\theta=\ee^{-\ii\theta\hat S_y}.
\end{aligned}
\label{eq:scs}
\end{equation}
Collective rotations act as identical rotations of each atom, and $\ket{j,j}=\ket{e}^{\otimes N}$, so that, up to an irrelevant global phase,
\begin{equation}
\begin{aligned}
    \ket{\Omega} &= \big( \cos\tfrac{\theta}{2}\ket{e}+\ee^{\ii\phi}\sin\tfrac{\theta}{2}\ket{g} \big)^{\otimes N}\\
    &= \big( \sqrt{z}\,\ket{e}+\ee^{\ii\phi}\sqrt{1-z}\,\ket{g} \big)^{\otimes N},
\end{aligned}
\label{eq:scs_product}
\end{equation}
where $z=\cos^2(\theta/2)$ is the excitation probability of each atom and $\phi$ the relative phase between $\ket{e}$ and $\ket{g}$. We denote by
\begin{equation}
    \Pi(z,\phi) \equiv \ket{\Omega}\!\bra{\Omega}, \qquad
    \overline\Pi(z) \equiv \int_0^{2\pi}\frac{\dd\phi}{2\pi}\,\Pi(z,\phi),
    \label{eq:Pibar}
\end{equation}
the projector onto the SCS and its azimuthal average.

Spin coherent states are not orthogonal and form an overcomplete set, $(2j+1)/(4\pi) \int\dd\Omega\,\ket{\Omega}\!\bra{\Omega}=\hat{\mathbb{I}}$,
where $\hat{\mathbb{I}}$ is the identity on the symmetric subspace. Every state of the collective spin admits a $P$ representation~\cite{Arecchi1972}, the spin analogue of the Glauber--Sudarshan representation,
\begin{equation}
    \hat\rho(\tau)=\int\dd\Omega\,P(\Omega,\tau)\ket{\Omega}\!\bra{\Omega},
    \label{eq:P_rep}
\end{equation}
normalized as $\int\dd\Omega\,P(\Omega,\tau)=1$, with $\dd\Omega=\sin\theta\,\dd\theta\,\dd\phi$. In general $P$ is not a probability distribution. If a non-negative $P$ exists, however, Eqs.~\eqref{eq:scs_product} and~\eqref{eq:P_rep} express $\hat\rho(\tau)$ as a convex combination of product states in which all atoms share the same state, so that $\hat\rho(\tau)$ is fully separable.

The $U(1)$ symmetry simplifies the search for such a $P$: since $\hat\rho(\tau)$ is invariant under $\hat U_\phi$, we can look for a $P$ independent of $\phi$ (see Sec.~\ref{sec:phasespace}) and write it as $P(\Omega,\tau)=\mathcal P(z,\tau)/(4\pi)$. Since $\dd\Omega=2\,\dd z\,\dd\phi$, Eq.~\eqref{eq:P_rep} becomes
\begin{equation}
    \hat{\rho}(\tau) = \int_0^1 \dd z\,\mathcal{P}(z,\tau)\,\overline\Pi(z),
    \qquad \int_0^1\dd z\,\mathcal P(z,\tau)=1,
    \label{eq:P_phase}
\end{equation}
and $\hat\rho(\tau)$ is fully separable if a solution $\mathcal P(z,\tau)\ge0$ exists.

\section{Main result}\label{sec:theorem}

In this section we show that Eq.~\eqref{eq:P_phase} holds with $\mathcal P(z,\tau)$ the probability density of a one-dimensional diffusion process. Let $z(\tau)\in[0,1]$ be the solution of the It\^o stochastic differential equation (derived in Sec.~\ref{sec:phasespace})
\begin{equation}
    \dd z = A(z)\,\dd\tau + \sqrt{D(z)}\,\dd W, \qquad z(0)=1,
    \label{eq:zSDE}
\end{equation}
with drift and diffusion coefficients
\begin{equation}
    A(z) = -z\bigl[N(1-z)+z\bigr], \qquad D(z) = 2z^2(1-z),
    \label{eq:AD}
\end{equation}
where $W(\tau)$ is a standard Wiener process.
At $z=1$ the diffusion vanishes and the drift $A(1)=-1$ points inwards, while at $z=0$ both coefficients vanish, so that $z(\tau)\in(0,1)$ for all $\tau>0$.

\begin{figure}[t]
  \includegraphics[width=\columnwidth]{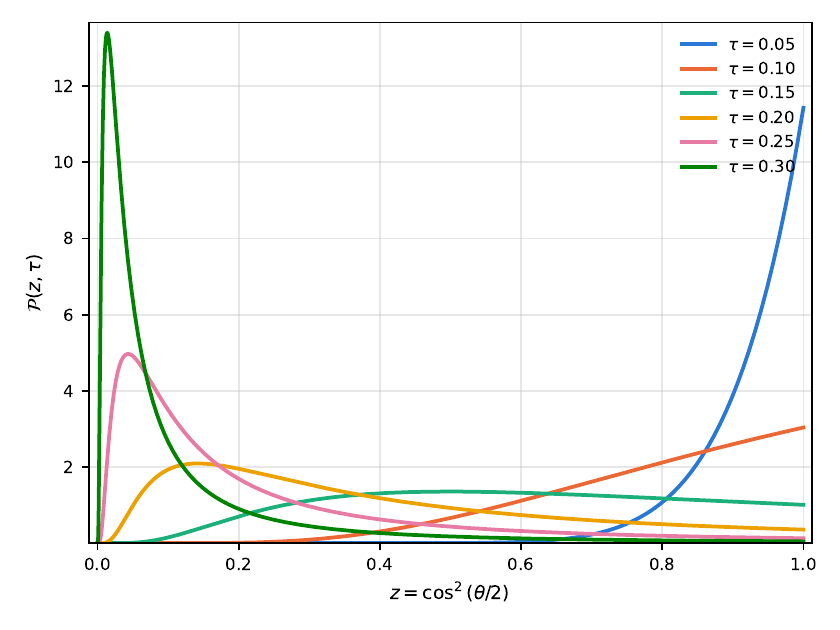}
  \caption{Probability density $\mathcal P(z,\tau)$ of the excitation
  probability per atom $z(\tau)$, Eq.~\eqref{eq:fpe}, for $N=20$ at different dimensionless times $\tau$.}
  \label{fig:Pz}
\end{figure}

\begin{theorem}
For every $N\ge1$ and every $\tau\ge0$, the solution of the master equation~\eqref{eq:master} with initial condition~\eqref{eq:initial_state} is
\begin{equation}
    \hat\rho(\tau) = \int_0^1\dd z\, \mathcal P(z,\tau)\,\overline\Pi(z),
    \label{eq:theorem}
\end{equation}
where $\mathcal P(z,\tau)$ is the probability density of $z(\tau)$ (Fig.~\ref{fig:Pz}).
Since $\mathcal P(z,\tau)\ge0$ by definition and the spin coherent states are product states, $\hat\rho(\tau)$ is fully separable.
\end{theorem}

\begin{proof}
The probability density $\mathcal P(z,\tau)$ obeys the Fokker--Planck equation
\begin{equation}
\begin{aligned}
\partial_\tau \mathcal P &= -\partial_z\bigl(A\,\mathcal P\bigr)
+ \tfrac{1}{2}\,\partial_z^2\bigl(D\,\mathcal P\bigr),\\
    \mathcal P(z,0) &= \delta(z-1),
\end{aligned}
\label{eq:fpe}
\end{equation}
where $\delta(z-1)$ denotes the point mass at $z=1$.
Let $\tilde\rho(\tau)$ denote the right-hand side of Eq.~\eqref{eq:theorem}, i.e., $\tilde\rho(\tau)=\EE\bigl[\,\overline\Pi(z(\tau))\bigr]$. An integration by parts in Eq.~\eqref{eq:fpe}, whose boundary terms vanish since, for $\tau>0$, both $D\mathcal P$ and the probability current $A\mathcal P-\tfrac12\partial_z(D\mathcal P)$ vanish at $z=0$ and $z=1$, gives
\begin{equation}
\frac{\dd \tilde\rho}{\dd \tau}
=\int_0^1\dd z\,\mathcal P(z,\tau)\,\mathcal F\bigl[\,\overline\Pi(z)\bigr],
\label{eq:dynkin}
\end{equation}
where $\mathcal F\equiv A(z)\,\partial_z+\tfrac12 D(z)\,\partial_z^2$ is the generator of the diffusion~\eqref{eq:zSDE}, i.e., the adjoint of the Fokker--Planck operator in Eq.~\eqref{eq:fpe}. Computing the $z$-derivatives of $\Pi(z,\phi)$ and comparing the action of $\mathcal F$ with that of the Lindblad superoperator~\eqref{eq:master}, one finds, for $0<z<1$,
\begin{equation}
\label{eq:kf:residue}
\mathcal L\big[\Pi(z,\phi)\big]
=
\mathcal F\big[\Pi(z,\phi)\big]
+
\frac{2z-1}{4(1-z)}\;\partial_\phi^2\,\Pi(z,\phi).
\end{equation}
The two generators differ by a second derivative in $\phi$ with a $\phi$-independent coefficient, which integrates to zero over the phase because $\Pi$ is $2\pi$-periodic in $\phi$. Hence $\mathcal F[\,\overline\Pi(z)]=\mathcal L[\,\overline\Pi(z)]$ for $0<z<1$, and by continuity on $[0,1]$. By linearity of $\mathcal L$,
\begin{equation}
\frac{\dd \tilde\rho}{\dd \tau}
=\int_0^1 \dd z\,\mathcal P(z,\tau)\,\mathcal L\big[\,\overline{\Pi}(z)\big]
=\mathcal L\big[\tilde\rho(\tau)\big].
\end{equation}
Since $\tilde\rho(0)=\overline\Pi(1)=\hat\rho(0)$, $\tilde\rho$ and $\hat\rho$ solve the same linear equation with the same initial condition; by uniqueness, $\tilde\rho(\tau)=\hat\rho(\tau)$ for all $\tau\ge0$.
\end{proof}

\section{Discussion}\label{sec:discussion}

The proof of Sec.~\ref{sec:theorem} only requires the identity~\eqref{eq:kf:residue}. In Sec.~\ref{sec:phasespace} we show how the diffusion~\eqref{eq:zSDE} arises from an exact phase-space description of the master equation. In Sec.~\ref{sec:moments} we relate the populations $p_m(\tau)$ of Eq.~\eqref{eq:dicke_diagonal} to the moments of the stochastic variable $z(\tau)$ of Eq.~\eqref{eq:zSDE}, explicitly solving the Hausdorff moment problem of Ref.~\cite{Bassler2025}.

\subsection{Phase-space origin of the diffusion}\label{sec:phasespace}
 
We describe the master equation with the Cahill--Glauber $k$-ordered quasiprobability distributions~\cite{CahillGlauber,CahillGlauberII}, where $k\in[-1,1]$ and $k=1,0,-1$ correspond to the Glauber--Sudarshan $P$, Wigner $W$, and Husimi $Q$ functions. To derive their evolution equation, we use the Schwinger representation $\hat S_i=\frac12\hat{\bm a}^\dagger\sigma_i\hat{\bm a}$ of the collective spin, restricted to the subspace with $\hat{\bm a}^\dagger\hat{\bm a}=2j$, where $\sigma_i$ are the Pauli matrices and $\hat{\bm a}=(\hat a_1,\hat a_2)^T$ are two bosonic modes associated with $\ket{e}$ and $\ket{g}$. We parametrize the phase-space variables $\bm\alpha=(\alpha_1,\alpha_2)^T$ in spherical coordinates,
\begin{equation}
\begin{aligned}
    \bm\alpha &= \sqrt{2r}\,\ee^{\ii\psi}\bigl(\cos\tfrac{\theta}{2},\,\ee^{\ii\phi}\sin\tfrac{\theta}{2}\bigr)^T,\\
    \bs &= r\,(\sin\theta\cos\phi,\,\sin\theta\sin\phi,\,\cos\theta),
\end{aligned}
\label{eq:spherical}
\end{equation}
where $\psi$ is a global phase and $s_i=\frac12\bm\alpha^\dagger\sigma_i\bm\alpha$ is the same phase-space representation of $\hat S_i$ for every $k$. Under the Schwinger map, the coherent state $\ket{\bm\alpha}$ projected onto the subspace with $2j$ bosons is proportional to the SCS $\ket{\theta,\phi}$ of Eq.~\eqref{eq:scs_product}. Following Ref.~\cite{Polkovnikov}, the chain rule turns the left and right $k$-ordered bosonic Bopp operators, acting on distributions independent of $\psi$, into the spin Bopp operators representing $\hat S_i\hat\rho$ and $\hat\rho\hat S_i$ (see also Ref.~\cite{ZuecoCalvo}),
\begin{equation}
\label{eq:kf:boppspin}
\begin{aligned}
\hat{L}_i/\hat{R}_i
={}&s_i-\frac{k}{2}\,r\,\partial_i
\mp\frac{\ii}{2}\,(\bs\times\bnab)_i
\\
&-\frac{1-k^2}{8}
\Big[
\partial_i+(\bs\cdot\bnab)\partial_i
-\tfrac{1}{2}\,s_i\nabla^2
\Big],
\end{aligned}
\end{equation}
where the upper (lower) sign refers to $\hat L_i$ ($\hat R_i$) and ${\partial_i\equiv\partial/\partial s_i}$.

Integrating out $\psi$, we reduce the dimension of the phase space: $\mathbb{C}^2 \simeq \mathbb{R}^4 \to \mathbb{R}^3$. The measure becomes $\dd^2\alpha_1\,\dd^2\alpha_2\to\pi r\sin\theta\,\dd r\,\dd\theta\,\dd\phi$ and we denote by $F^{(k)}(r,\theta,\phi,\tau)$ the density with respect to $\dd r\,\dd\theta\,\dd\phi$. For $k=\pm1$ the last term in Eq.~\eqref{eq:kf:boppspin} vanishes, the Bopp operators are of first order, and the master equation~\eqref{eq:master} maps exactly onto
\begin{equation}
\begin{aligned}
\partial_\tau F^{(k)}
={}&-\partial_\theta\bigl(A_\theta F^{(k)}\bigr)
+\tfrac12\,\partial_\theta^2\bigl(D_{\theta\theta}F^{(k)}\bigr)\\
&+\tfrac12\,\partial_\phi^2\bigl(D_{\phi\phi}F^{(k)}\bigr)
+\partial_\theta\partial_r\bigl(D_{\theta r}F^{(k)}\bigr),
\end{aligned}
\label{eq:kf:FPE_Full}
\end{equation}
with the nonzero coefficients
\begin{equation}
\begin{alignedat}{2}
A_\theta &= r\sin\theta+\frac{\cos\theta+k}{2\sin\theta}, &\quad
D_{\theta\theta} &= 1+k\cos\theta,\\
D_{\phi\phi} &= \frac{\cos\theta\,(\cos\theta+k)}{\sin^2\theta}, &\quad
D_{\theta r} &= \tfrac{k}{2}\,r\sin\theta .
\end{alignedat}
\label{eq:kf:coeff}
\end{equation}

Equation~\eqref{eq:kf:FPE_Full} shows why no ordering directly yields an exact diffusion. For $|k|<1$ the exact equation also contains higher-order derivatives; for $k=0$, neglecting all terms of order $1/j$, which include them, gives the truncated Wigner approximation valid for $j\gg1$~\cite{Polkovnikov}, whose equation has the form~\eqref{eq:kf:FPE_Full} with a positive-semidefinite diffusion matrix. For $k=\pm1$ the equation is exact, but $D_{\phi\phi}$ changes sign at $\theta=\pi/2$ and the mixed term $D_{\theta r}$ couples the polar angle to the radial variable $r$.

Both obstacles are removed by the symmetries of the initial state. Under the Schwinger map, $\ket{j,j}$ is the two-mode Fock state with $2j=N$ bosons in mode $1$ and none in mode $2$. Using Eq.~\eqref{eq:spherical}, the joint quasiprobability distributions of the two bosonic modes can be mapped to an initial distribution uniform in $\phi$ for every $k$. Moreover, for $k=\pm1$ it factorizes exactly in $r$ and $\theta$:
\begin{equation}
   F^{(k)}(r,\theta,\phi,0)=\chi^{(k)}(r)\,f^{(k)}(\theta,0)/(2\pi). 
\end{equation}
Defining $R\equiv\avg{r}_k=j+(1-k)/2$, the radial marginal satisfies
\begin{equation}
\tfrac{k}{2}\,\partial_r\bigl(r\chi^{(k)}\bigr)=(r-R)\,\chi^{(k)}.
\label{eq:radial}
\end{equation}
For $k=1$, $\chi^{(1)}$ and $f^{(1)}(\theta,0)=\delta(\theta)$ are generalized functions, as the $P$ function of a Fock state is singular. Since the coefficients~\eqref{eq:kf:coeff} do not depend on $\phi$, inserting $F^{(k)}(r,\theta,\phi,\tau)=\chi^{(k)}(r)\,f^{(k)}(\theta,\tau)/(2\pi)$ into Eq.~\eqref{eq:kf:FPE_Full} shows that this form is preserved at all times. The $\phi$-diffusion vanishes and, by Eq.~\eqref{eq:radial}, the mixed term reduces to $(r-R)\,\chi^{(k)}\,\partial_\theta\bigl(\sin\theta\,f^{(k)}\bigr)/(2\pi)$, which cancels the $r$-dependence of the polar drift and replaces $r\sin\theta$ by $R\sin\theta$. The polar distribution therefore obeys
\begin{equation}
\partial_\tau f^{(k)}
=-\partial_\theta\bigl(A_\theta\big|_{r=R}\,f^{(k)}\bigr)
+\tfrac12\,\partial_\theta^2\bigl(D_{\theta\theta}f^{(k)}\bigr).
\label{eq:FPE_theta}
\end{equation}

For $k=1$ one has $R=j=N/2$ and $f^{(1)}(\theta,0)=\delta(\theta)$, and the change of variables $z=\cos^2(\theta/2)$, with $f^{(1)}(\theta,\tau)\,\dd\theta=\mathcal P(z,\tau) \dd z$, maps Eq.~\eqref{eq:FPE_theta} onto the Fokker--Planck equation~\eqref{eq:fpe}, whose It\^o form is Eq.~\eqref{eq:zSDE}. Up to the reflection $\theta\to\pi-\theta$, Eq.~\eqref{eq:FPE_theta} with $k=1$ is the exact diffusion equation of Ref.~\cite{Narducci1974}, derived there directly for the phase-averaged $P$ function.

This phase-space construction explains the identity~\eqref{eq:kf:residue}. Without the phase average, the factorized form $F^{(1)}=\chi^{(1)}(r)\,g(\theta,\phi,\tau)$ is still preserved, and $g(\theta,\phi,\tau)=P(\Omega,\tau)\sin \theta$ obeys the exact evolution equation of the $P$ function on the sphere, whose generator, written in the variables $(z,\phi)$, is $\mathcal F+\frac12 D_{\phi\phi}\,\partial_\phi^2$, with $D_{\phi\phi}=(2z-1)/[2(1-z)]$ [Eq.~\eqref{eq:kf:coeff}, $k=1$]. Equation~\eqref{eq:kf:residue} states precisely that this generator reproduces the action of $\mathcal L$ on the coherent-state projectors. Since $D_{\phi\phi}<0$ for $z<1/2$, this equation does not describe a diffusion process. The phase average, allowed by the $U(1)$ symmetry, removes the $\phi$-diffusion and leaves the polar marginal~\eqref{eq:FPE_theta}, whose generator $\mathcal F$ has non-negative diffusion.

\subsection{Moments of \texorpdfstring{$z$}{z} and Dicke populations}\label{sec:moments}

Bassler demonstrated that Dicke superradiance does not produce entanglement by proving the existence, at all times, of a probability distribution of the excitation probability $z\in[0,1]$ whose first $N$ moments are fixed by the Dicke populations~\cite{Bassler2025}. The density $\mathcal P(z,\tau)$ solving Eq.~\eqref{eq:fpe} provides such a distribution explicitly. In this subsection we connect Eq.~\eqref{eq:zSDE} to the truncated Hausdorff moment problem studied by Bassler, obtaining an alternative proof of the theorem of Sec.~\ref{sec:theorem}.

Inserting Eq.~\eqref{eq:dicke_diagonal} into Eq.~\eqref{eq:master} shows that the Dicke populations $p_m(\tau)$ obey the pure death process
\begin{equation}
    \dot p_m = \Gamma_{m+1}\,p_{m+1}-\Gamma_m\,p_m,
    \label{eq:death_process}
\end{equation}
with rates $\Gamma_m=(j+m)(j-m+1)$ and $p_m(0)=\delta_{m,j}$.

Let $\mu_n(\tau)\equiv\EE[z^n]=\int_0^1\,z^n\,\mathcal P(z,\tau)\dd z$, $n=0,\dots,N$, be the moments of the process~\eqref{eq:zSDE}. Applying It\^o's formula to $z^n$ and averaging gives, with $\mu_n(0)=1$ for all $n$, the closed, upper-triangular hierarchy
\begin{equation}
    \dot\mu_n = -n(N-n+1)\,\mu_n + n(N-n)\,\mu_{n+1}.
    \label{eq:z_moment_hierarchy}
\end{equation}
Since $z$ is the probability that a single atom is excited, $\mu_n$ is the probability that $n$ given atoms out of $N$ are all excited. When $j+m$ atoms are excited, this happens with probability $\binom{j+m}{n}/\binom{N}{n}$, so that
\begin{equation}
\begin{aligned}
    \mu_n \equiv \mathbb{E}[z^n]
    &= \binom{N}{n}^{-1}
    \sum_{m=-j}^{j}\binom{j+m}{n}\,\tilde p_m, \\
    \mu_n(0)=1
    &\implies \tilde p_m(0)=\delta_{m,j},
\end{aligned}
\end{equation}
where $\tilde p_m$ is defined as the probability that exactly $j+m$ of the $N$ atoms are excited,
\begin{align}
    \tilde p_m &\equiv \EE\bigl[|\braket{j,m|\Omega}|^2\bigr] \notag\\
    &= \binom{N}{j+m}\EE\bigl[z^{j+m}(1-z)^{j-m}\bigr] \notag\\
    &= \binom{N}{j+m}\sum_{\ell=0}^{j-m}(-1)^{\ell}\binom{j-m}{\ell}\,\mu_{j+m+\ell}.
    \label{eq:binomial_transform}
\end{align}
The moments $\mu_n$ are thus in one-to-one correspondence with the probabilities $\tilde p_m$. Taking the time derivative of Eq.~\eqref{eq:binomial_transform} and using Eq.~\eqref{eq:z_moment_hierarchy}, one finds that the $\tilde p_m$ obey the death process~\eqref{eq:death_process}; with the initial condition $\tilde p_m(0)=\delta_{m,j}$, this proves that $\tilde p_m(\tau)=p_m(\tau)$.

Expanding Eq.~\eqref{eq:scs_product} in the Dicke basis shows that the phase average removes all coherences,
\begin{equation}
    \overline\Pi(z)=\sum_{m}\binom{N}{j+m}z^{j+m}(1-z)^{j-m}\ket{j,m}\!\bra{j,m},
    \label{eq:Pibar_dicke}
\end{equation}
so that, by Eq.~\eqref{eq:dicke_diagonal},
\begin{equation}
    \int_0^1\dd z\,\mathcal P(z,\tau)\,\overline\Pi(z)=\sum_m\tilde p_m\ket{j,m}\!\bra{j,m}=\hat\rho(\tau).
\end{equation}
This is an alternative proof of Eq.~\eqref{eq:theorem}. We conclude that the Dicke populations are binomial distributions averaged over the trajectories of Eq.~\eqref{eq:zSDE}.

\section{Conclusions}\label{sec:conclusions}

We have shown that, starting from the fully excited state, the atomic density matrix of the Dicke cascade is fully separable for every $N$ and at all times. It is a statistical mixture of spin coherent states, i.e., of product states in which all atoms share the same single-atom state, so purely collective decay does not generate entanglement among the atoms. The result rests on the two symmetries of the model. Conservation of the total spin confines the dynamics to the symmetric subspace, where the master equation maps exactly onto a second-order equation for the $P$ function over SCS. This equation does not describe a diffusion, because its azimuthal diffusion coefficient is negative for $z<1/2$. The $U(1)$ symmetry allows us to replace $P$ by its azimuthal average, which removes the azimuthal diffusion and leaves a one-dimensional equation for $z$ with non-negative diffusion. Its solution $\mathcal P(z,\tau)$ is therefore a genuine probability density, that of the stochastic process~\eqref{eq:zSDE}, and provides an explicit separable decomposition of $\hat\rho(\tau)$. More generally, our analysis shows how the symmetries of a problem can turn an exact phase-space equation that is not a diffusion into one that is, without approximation.

\begin{acknowledgments}
This work was carried out during an Erasmus+ exchange from the University of Padova to the University of Strasbourg; I am grateful to the Erasmus+ program for this opportunity and I thank CESQ at the University of Strasbourg for hospitality. I am grateful to my supervisors Tom Bienaim\'e and J\'er\^ome Dubail for their guidance, and I particularly thank J.~Dubail for his careful and critical reading of this manuscript. I also thank Johannes Schachenmayer, Luiz O.~R. Solak and Andreas Buchleitner for discussions and for their interest in this work. I acknowledge the use of the AI tool Claude (Anthropic), which assisted me with writing simulations and checking computations.

This work was supported by the France 2030 project ``Dyn1D'' of PEPR Quantique (ANR-23-PETQ-0001). It was also supported by the Interdisciplinary Thematic Institute QMat, which is part of the ITI 2021--2028 program of the University of Strasbourg, CNRS and Inserm, supported by IdEx Unistra (ANR-10-IDEX-0002), and by the SFRI STRAT'US project (ANR-20-SFRI-0012) and EUR QMAT (ANR-17-EURE-0024) under the framework of the French Investments for the Future Program. I also acknowledge support from the Seed Money project ``Open quantum many-body systems'' (SM24\_\#9\_S\_BFK) funded by Eucor -- The European Campus.
\end{acknowledgments}

\makeatletter\@booleantrue\raggedcolumn@sw\makeatother
\bibliographystyle{apsrev4-2}
\bibliography{references}

\end{document}